\documentclass[prd,twocolumn,showpacs,floats,letterpaper,nofootinbib]{revtex4}
\usepackage{amssymb,amsmath,amsfonts}
\usepackage{color}
\usepackage{graphicx}

\newcommand{\s}{{\rm s}}

\newcommand{\pp}{{\phi\phi}}

\newcommand{\bp}{{\cal C}}

\newcommand{\tot}{{\rm t}}
\newcommand{\shell}{{\rm s}}

\newcommand{\cmb}{{\Theta}}

\newcommand{\len}{\phi}

\newcommand{\bea}{\begin{eqnarray}}
\newcommand{\eea}{\end{eqnarray}}
\newcommand{\bean}{\begin{eqnarray*}}
\newcommand{\eean}{\end{eqnarray*}}

\newcommand{\vecl}{{\bf \hat{l}}}
\newcommand{\vecla}{{{\bf l}_1}}
\newcommand{\veclb}{{{\bf l}_2}}
\newcommand{\veclc}{{{\bf l}_3}}
\newcommand{\vecld}{{{\bf l}_4}}

\newcommand{\bfl}{{\mathbf{l}}}
\newcommand{\dirac}{{\rm D}}

\newcommand\lsim{\mathrel{\rlap{\lower4pt\hbox{\hskip1pt$\sim$}}
    \raise1pt\hbox{$<$}}}
\newcommand\gsim{\mathrel{\rlap{\lower4pt\hbox{\hskip1pt$\sim$}}
    \raise1pt\hbox{$>$}}}

\newcommand{\joe}[1]{}

\begin{document}

\title{The Impact of Secondary non-Gaussianities in the CMB on \\ Cosmological Parameter Estimation} 
\author{Joseph Smidt,  Shahab Joudaki, Alexandre Amblard, Paolo Serra, and Asantha  Cooray} 
\affiliation{Center for Cosmology,
Department of Physics  and  Astronomy, 
University  of California, Irvine, CA 92697}

\begin{abstract} 
We consider corrections to the underlying cosmology due to secondary
contributions from weak gravitational lensing, the integrated Sachs-Wolfe effect, and the
Sunyaev-Zel'dovich effect contained in the trispectrum. We incorporate these additional contributions to the covariance of a binned angular power spectrum of temperature anisotropies in the analysis of current and prospective data sets. Although recent experiments such as ACBAR and CBI are not particularly sensitive to these additional non-Gaussian effects, the interpretation of Planck and CMBPol anisotropy spectra will
require an accounting of non-Gaussian covariance leading to a degradation in cosmological parameter estimates by up to $20\%$ and $30\%$, respectively. 
\end{abstract}


\maketitle

\section{Introduction} 

The cosmic microwave background's (CMB) ability to constrain the cosmological parameters has driven a large variety of CMB experiments.  The high quality measurements of the temperature and polarization of the CMB are compatible with a flat universe with nearly scale invariant fluctuations. The sensitivity to small angular scales of recent experiments such as WMAP5~\cite{WMAP:2008ie}, ACBAR~\cite{Acbar:2008ay} and CBI~\cite{CBI:2009ah} give us a better understanding of anisotropies from local large scale structure (LSS) and from non-linear effects such as weak
gravitational lensing. The non-Gaussianity of these fluctuations can have an effect in the estimation of the cosmological parameters, especially for future experiments. However, possible non-Gaussian effects have not been considered in previous analyses, under the assumption that they are negligible.

Non-Gaussianities show up as contributions to the
four-point correlation function, or trispectrum in Fourier space, of the
CMB temperature fluctuations~\cite{Hu:2000ee,Zaldarriaga:2000ud}.  
The four point correlations quantify the sample variance and covariance
of the two point correlation of power spectrum measurements~\cite{Scoccimarro:1999kp,Cooray:2000ry}.  To adequately understand the
statistical measurements of CMB anisotropy fluctuations a proper understanding of the four point contributions is needed.
Given the high precision level of cosmological parameter
measurements by current and future surveys, a careful consideration must be
attached to understanding the presence of non-Gaussian signals at the
four point level. The goal of this paper is  to understand to
what extent these non-Gaussianities affect constraints on the cosmological parameters

As explored in previous work (e.g.~\cite{Cooray:1999kg,Spergel:1999xn,Zaldarriaga:1998te,Peiris:2000kb}),
one of the most important non-linear contributions to CMB temperature
fluctuations results from weak gravitational lensing. 
Contributions to the four point level arise from the non-linear
mode-coupling nature of the lensing effect, as well as correlations between
weak lensing angular deflections and secondary effects that trace the
same large scale structure. The trispectrum due to lensing alone is
studied in~\cite{Zaldarriaga:2000ud}, and further considered under an all-sky
formulation in~\cite{Hu:2000ee}.  

In our analysis we include contributions to the trispectrum from three
well known secondary sources. First, we consider the contribution of the
trispectrum to the power spectrum covariance due to lensing.  We also
consider contributions from the integrated Sachs-Wolfe effect
(ISW;~\cite{Sachs:1967er}) and the Sunyaev-Zel'dovich effect
(SZ;~\cite{Sunyaev:1980nv}), both which cross-correlate with lensing to produce non-Gaussianity. Each of these effects becomes non-negligible on small
angular scales and must be taken into account for a more complete treatment of cosmological parameter constraints.

In a situation without instrumental noise, the lensing contribution to the trispectrum can increase the values along the diagonal of the covariance matrix on the order of about $10\%$~\cite{Cooray:2002fy}. The SZ contributions are greater than the lensing by as much as an order of magnitude. However, with noise added follows the possibility that these effects on the trispectrum become negligible. Most experiments thus far have been too noisy at small angular scales to observe these non-Gaussian effects on the cosmological parameters. We explore how firmly this holds for the ACBAR and CBI data sets when combined with WMAP five-year data as well for future Planck and CMBPol measurements.  

Moreover, we consider the weak lensing scaling parameter $A_L$ in our analysis.
This parameter scales the power spectrum of the CMB lensing potential $C_l^{\phi \phi}$ such that
$A_L = 0$ corresponds to an unlensed scenario, whereas $A_L = 1$
renders the expected lensed result. It was first reported that in an
analysis of the ACBAR data set this parameter is incompatible
with the expected  value of $A_L = 1$ at $2.5\sigma$~\cite{Calabrese:2008rt}.
ACBAR has since updated their data set with a new calibration, 
and the ACBAR team finds that the amplitude is consistent with the expected value of unity \cite{Reichardt:2008ay}.
However, given that this parameter is sensitive to physics at high multipoles, such as that
encapsulated in the trispectrum, we also include $A_L$ as an extra parameter in our
calculations.

\begin{figure}
\centering
\includegraphics[scale=0.47]{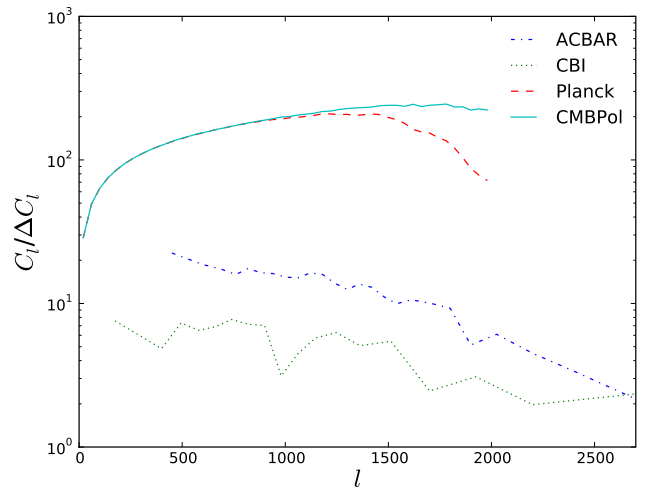}
\caption{The multipole-dependent  ratio $C_l/\Delta{C_l}$ for the considered experiments (ACBAR in dot-dashed, CBI in dotted, Planck in dashed, and CMBPol in solid).}
\label{fig:sig_noise}
\end{figure}

Our analysis of the ACBAR and CBI data sets, combined with WMAP 5-year data, show that secondary contributions in the trispectrum have a negligible effect on the uncertainties of the cosmological parameters. At most the constraints on the cosmological parameters change by no more than 10\%. We find that these modifications are negligible due to the increase in noise at large multipoles (as shown in Fig.~\ref{fig:sig_noise}). 

However, in a mock Planck data set with realistic noise contributions, the non-Gaussian effects coming from the trispectrum become more important, with error constraint degradations up to $20\%$ (as shown in Table \ref{tab:planck}). For the case of CMBPol this rises to $\lsim30\%$. For this reason, future experiments with at least Planck-level sensitivity at high $l$ should consider non-Gaussianities from the trispectrum in order to have proper error bar estimates. 

Lastly, we show $A_L$ to be consistent with the expected value of unity, primarily due to the same reasons that Ref.~\cite{Reichardt:2008ay} found it to be consistent with the theoretically expected value. If we had used the original ACBAR dataset, prior to their most recent update, we find that
$A_L$ is inconsistent with unity at the same confidence as found by Ref.~\cite{Calabrese:2008rt}.
We also find that this parameter remains consistent with unity when both trispectrum effects and the running $dn_s/dlnk$ of the spectral index $n_s$  are included in the analysis.

In \S \ref{sec:analysis}, we review the calculation for the trispectrum contribution to the
covariance matrix. In \S \ref{sec:results}, we
calculate the impact of the trispectrum on the cosmological parameters for current and future experiments considering a set of parameter configurations.  The fiducial cosmologies considered can be found in Table~\ref{tab:fiducial}. In \S \ref{sec:summary}, we conclude with a summary. 

\section{Analysis}
\label{sec:analysis}

\subsection{Lensing Trispectrum}
We begin with a review of how to calculate the various contributions of the trispectrum to the covariance matrix. For a comprehensive introduction and derivation of these
equations, see~\cite{Cooray:2002fy}. 

In the flat sky approximation, the power spectrum and trispectrum are defined via:
\begin{eqnarray}
\left< \cmb^\tot(\bfl_1)\cmb^\tot(\bfl_2)\right> &=&
        (2\pi)^2 \delta_\dirac(\bfl_{12}) C_l^\cmb\,,\nonumber\\
\left< \cmb^\tot(\bfl_1) \ldots
       \cmb^\tot(\bfl_4)\right>_c &=& (2\pi)^2
\delta_\dirac(\bfl_{1234})
        T^\cmb(\bfl_1,\bfl_2,\bfl_3,\bfl_4)\,. \nonumber \\
\end{eqnarray}

In computing these quantities  the band powers are obtained, as 

\begin{equation}
\bp_i = \int_{\shell_i}
{d^2 l W_i\over{A_{\shell_i}}}
\frac{l^2}{2\pi} \cmb({\bf l}) \cmb(-\bf l) \, ,
\end{equation}

where $A_{\shell_{i}} \equiv A_\shell(l_i) = \int d^2 l W_i$ is the area of 2D shell in multipole space, $W_i$ is the $i^{\rm th}$ window function and the subscript $\shell_i$ stands for the $i^{\rm th}$ shell in multipole space over which we are integrating.  The signal covariance matrix $C_{ij}$ is given by

\begin{eqnarray}
C_{ij} &=& {1 \over A} \left[ {(2\pi)^2 \over A_{\shell_i}} 2 \bp_i^2
+ T^\cmb_{ij}\right]\,,\\
T^\cmb_{ij}&=&
\int {d^2 l_i W_i \over A_{\shell_i}}
\int {d^2 l_j W_j \over A_{\shell_j}} {l_i^2 l_j^2 \over (2\pi)^2}
T^\cmb(\bfl_i,-\bfl_i,\bfl_j,-\bfl_j)\,,
\label{eqn:variance}
\end{eqnarray}

where $A$ is the survey area in steradians. The first term of the covariance matrix $C_{ij}$ is the Gaussian contribution to the
sample variance and includes, in addition to the primary component, contributions through lensing and secondary effects. The second term $T_{ij}$
represents the non-Gaussian contributions contained in the trispectrum.

The surveys we consider also include instrumental noise. The noise is incorporated into our Gaussian variance as an additional contribution to the power spectrum  
\begin{equation}
C_l^\tot = C_l^\cmb+N_l\,
\end{equation}

where $N_l$ is the power spectrum of the detector and other sources of noise
introduced by the experiment. These noise contributions are included in respective experiment's publicly released data and significantly impact the covariance matrix, in particular on small angular scales.

For the power spectrum covariance, we are interested in
the case where $\veclb = -\vecla$  with $|\vecla|=l_i$, and
$\vecld = -\veclc$ with $|\veclc|=l_j$. These conditions render parallelograms for
the trispectrum configuration in Fourier space.

To account for the lensing contribution to the trispectrum, we must compute
\begin{eqnarray}
&& T^\cmb(\bfl_i,-\bfl_i,\bfl_j,-\bfl_j) = \nonumber \\
&&C_{l_i}^\cmb C_{l_i}^\cmb \left[
C_{|\vecl_i+\vecl_j|}^{\pp} \left[ (\vecl_i+\vecl_j)\cdot
\vecl_i\right]^2 +
C_{|\vecl_i-\vecl_j|}^{\pp} \left[ (\vecl_i-\vecl_j)\cdot
\vecl_i\right]^2 \right] \nonumber \\
&+&C_{l_j}^\cmb C_{l_j}^\cmb \left[
C_{|\vecl_i+\vecl_j|}^{\pp} \left[ (\vecl_i+\vecl_j)\cdot
\vecl_j\right]^2 +
C_{|\vecl_i-\vecl_j|}^{\pp} \left[ (\vecl_i-\vecl_j)\cdot
\vecl_j\right]^2 \right] \nonumber \\
&+&2C_{l_i}^\cmb C_{l_j}^\cmb \Big[
C_{|\vecl_i+\vecl_j|}^{\pp}  (\vecl_i+\vecl_j)\cdot
\vecl_i(\vecl_i+\vecl_j)\cdot \vecl_j \nonumber \\
&& \quad \quad -
C_{|\vecl_i-\vecl_j|}^{\pp} (\vecl_i-\vecl_j)\cdot \vecl_i
(\vecl_i-\vecl_j)\cdot \vecl_j \Big] \, ,
\label{eqn:lenTri}
\end{eqnarray}
where $C_{l_i}^\cmb$ is the TT power spectrum and $C_{l_i}^{\pp}$ is the lensing power spectrum. This equation includes all terms with no additional permutations. Similarly, for the lensing-secondary trispectrum effects, such as lensing-ISW and lensing-SZ, we calculate
\begin{eqnarray}
&& T^\cmb(\bfl_i,-\bfl_i,\bfl_j,-\bfl_j) = \nonumber \\
&&2 \left(\bfl_i \cdot \bfl_j\right)^2 \left[
\left(C_{l_i}^{\len\s}\right)^2 C_{l_j}^\cmb + 
\left(C_{l_j}^{\len\s}\right)^2 C_{l_i}^\cmb \right] \nonumber \\
&-& \left[\left[\bfl_i \cdot (\bfl_i+\bfl_j)\right]^2 
\left(C_{l_i}^{\len\s}\right)^2 
+\left[\bfl_j \cdot (\bfl_i+\bfl_j)\right]^2 
\left(C_{l_j}^{\len\s}\right)^2 \right]C_{|\bfl_i+\bfl_j|}^\cmb
\nonumber \\
&-& \left[\left[\bfl_i \cdot (\bfl_i-\bfl_j)\right]^2 
\left(C_{l_i}^{\len\s}\right)^2 
+ \left[\bfl_j \cdot (\bfl_i-\bfl_j)\right]^2 
\left(C_{l_j}^{\len\s}\right)^2\right] C_{|\bfl_i-\bfl_j|}^\cmb
\nonumber \\
&+& 2\left[\bfl_i \cdot (\bfl_j-\bfl_i)\right]
\left[\bfl_j \cdot (\bfl_j-\bfl_i)\right]
C_{l_i}^{\len\s}C_{l_j}^{\len\s} C_{|\bfl_j-\bfl_i|}^\cmb \nonumber \\
&-& 2\left[\bfl_i \cdot (\bfl_i+\bfl_j)\right]
\left[\bfl_j \cdot (\bfl_i+\bfl_j)\right]
C_{l_i}^{\len\s}C_{l_j}^{\len\s} C_{|\bfl_i+\bfl_j|}^\cmb \nonumber \\
\label{eqn:secTri}
\end{eqnarray}

where the $s$ is a place holder denoting either the ISW or SZ contribution.

\begin{figure}[!b]
\centering
\includegraphics[scale=0.43]{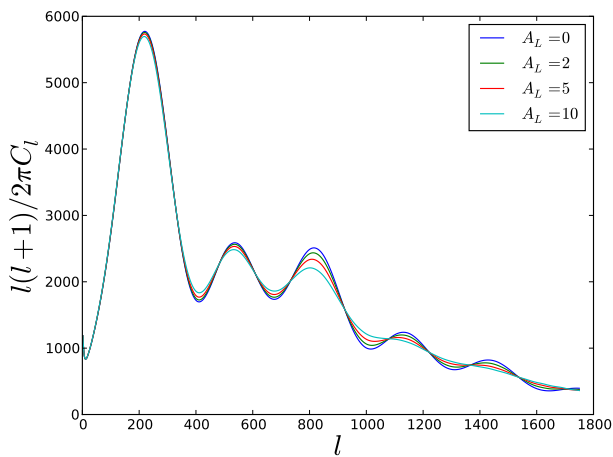}
\caption{The impact of varying the lensing scaling parameter on the lensed CMB temperature power spectrum, for $A_L$ = [0,2,5,10].}
\label{fig:AL}
\end{figure}

\subsection{SZ Trispectrum}

In addition to the lensing contributions to the trispectrum above, we consider contributions from the inverse Compton scattering of the CMB photons. The SZ contribution to the trispectrum is given by~\cite{Komatsu:2002wc,Cooray:2001wa}:

\begin{eqnarray}
 \nonumber
 T^\cmb_{ij}&=& g_\nu^4 \int_0^{z_{\rm max}} dz \frac{dV}{dz}
              \int_{M_{\rm min}}^{M_{\rm max}} dM \frac{dn(M,z)}{dM}\\
 \label{eq:tl}
	& &\times \left|\tilde{y}_i(M,z)\right|^2
	\left|\tilde{y}_{j}(M,z)\right|^2,
	\label{eqn:komSZ}
\end{eqnarray}

where $g_{\nu}$ is the spectral function of the SZ effect, $V(z)$ is the comoving volume of the universe integrated to a redshift of $z_{\rm max} = 4$, $M$ is the virial mass such that $[{\rm log}_{10}(M_{\rm min}), {\rm log}_{10}(M_{\rm max})] = [11,16]$, ${dn}/{dM}$ is the mass function of dark matter halos as rendered by~\cite{Jenkins:2000bv} utilizing the linear transfer function of~\cite{Eisenstein:1997jh}, and $\tilde{y}$ is the dimensionless two-dimensional Fourier transform of the projected Compton $y$-parameter, given via the Limber approximation \cite{Limber:1954rt} by:

\begin{equation}
 \label{eq:yl}
  \tilde{y}_l= \frac{4\pi r_{\rm s}}{l_{\rm s}^2}
  \int_0^\infty dx x^2 y_{\rm 3D}(x)
  \frac{\sin(lx/l_{\rm s})}{lx/l_{\rm s}},
\end{equation}

where the scaled radius $x=r/r_s$ and $l_s=d_A/r_s$ such that $d_A$ is the angular diameter distance and $r_s$ is the scale radius of the three-dimensional radial profile $y_{\rm 3D}$ of the Compton $y$-parameter. This  profile is a function of the gas density and temperature profiles as modeled in \cite{Komatsu:2001dn}. Hence, we incorporate the contributions obtained from the SZ effect along with those from lensing, lensing-ISW, and lensing-SZ effects to the covariance matrix in Eqn.~3. 

\subsection{The Weak Lensing Scaling Parameter $A_L$}

To first order in $\phi$, the weak lensing of the CMB anisotropy trispectrum can be expressed as the convolution of the power spectrum of the unlensed temperature $C_l$ and that of the weak lensing potential $Cl^{\phi \phi}$~\cite{Calabrese:2008rt,Lewis:2006fu,Zaldarriaga:1998ar}. The
magnitude of the lensing potential power spectrum can be parameterized by
the scaling parameter $A_L$, defined as
\begin{equation}
 \label{eq:wel}
 C_l^{\phi \phi} \rightarrow A_L C_l^{\phi \phi}.
\end{equation}

Thus, $A_L$ is a measure of the degree to which the expected amount of
lensing appears in the CMB, such that a theory with $A_L=0$ is
devoid of lensing, while $A_L=1$ renders a theory with the canonical
amount of lensing. Any inconsistency with unity represents an unexpected
amount of lensing that needs to be explained with new physics, such as
dark energy or modified gravity \cite{Calabrese:2008rt,Calabrese:2009tt}. The impact of
this scaling parameter on the lensed CMB temperature power spectrum can be
seen in Fig.~\ref{fig:AL}. Qualitatively, $A_L$ smoothes out the peaks in
the power spectrum and can therefore also be viewed as a smoothing
parameter in addition to its scaling property. Given that $A_L$ primarily
affects the temperature power spectrum on small angular scales, we also explore the possibility that
it deviates from unity  as secondary non-Gaussianities are accounted for in the analysis.

\section{Results}
\label{sec:results}

\subsection{Analysis Using ACBAR and CBI Data}

\begin{table}[htbp]
  \centering
  \begin{tabular}{@{} |c|c|c| @{}}
    \hline
     & ACBAR & CBI \\ 
    \hline
     $H_0$ & 72.0 & 73.2 \\ 
    $\Omega_b h^2$  & 0.02282 & 0.02291 \\ 
    $\Omega_c h^2$ & 0.1108 & 0.1069 \\ 
    $\tau$& 0.088 & 0.086 \\
    $n_s$ & 0.964 & 0.960 \\ 
    $\Delta^2_R$ & $2.41\times10^{-9}$ & $2.09\times10^{-9}$ \\ 
    \hline
  \end{tabular}
  \caption{Fiducial model used to generate $C_{l_i}^\cmb$ and $C^{\phi \phi}_{l_i}$ in CAMB for ACBAR and CBI.}
  \label{tab:fiducial}
\end{table}

\begin{figure}
\centering
\includegraphics[scale=0.47]{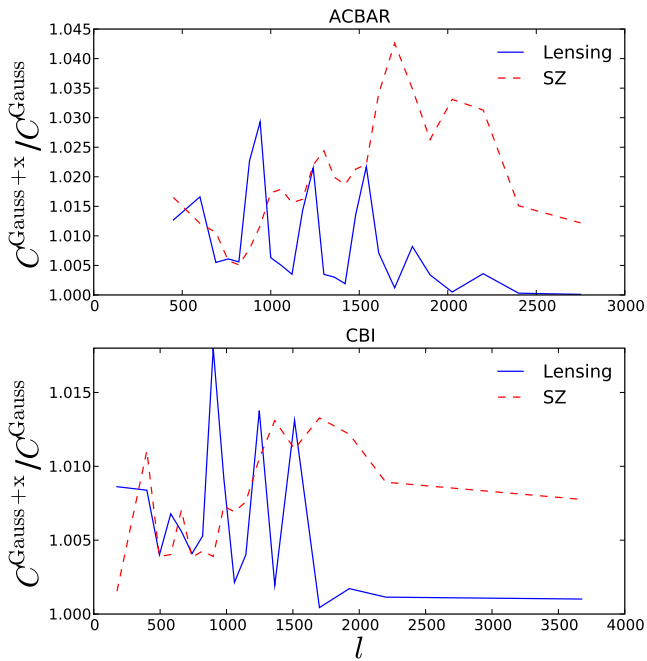}
\caption{The ratio of the diagonal of the covariance matrix plus trispectrum effects with the diagonal of the covariance matrix neglecting trispectrum contributions for ACBAR (top) and CBI (bottom). The $x$ in the
$C_l^{\mathrm{Gauss+x}}$ is the specific contribution being added. We see that
both the combined lensing effects and SZ alter the covariance matrix by at most a few
percent. The choppiness is due to two issues: the covariance matrices provided by ACBAR and CBI are not smooth and the window functions used to bin the data take on different shapes and are themselves not smooth.}
\label{fig:ratio}
\end{figure}

We compute $C_{l_i}^\cmb$ and $C^{\phi \phi}_{l_i}$, separately for each aforementioned experiment by utilizing the publicly available\footnote{\rm http://camb.info/} Fortran code CAMB~\cite{Lewis:1999bs} for the fiducial models found in Table~\ref{tab:fiducial}. These fiducial models were taken from NASA's Lambda website\footnote{\rm http://lambda.gsfc.nasa.gov/} which houses many best fit fiducial models derived from several different data sets. 
From these spectra, we use Eqns.~\ref{eqn:lenTri}--\ref{eqn:komSZ} to determine the trispectrum
contributions from lensing, lensing-ISW, lensing-SZ, and SZ effects.  We then add each corresponding trispectrum contribution to each experiment's
publicly available covariance matrix.

\begin{figure}
\centering
\includegraphics[scale=0.47]{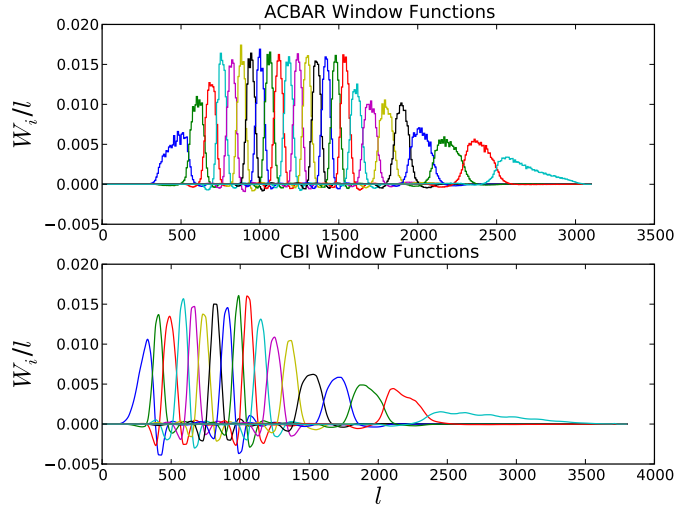}
\caption{The window functions used for our analysis of the ACBAR and CBI
data sets.}
\label{fig:windows}
\end{figure}

For our analysis, we use the same multipole binning as the aforementioned experiments, which are obtained from the window functions plotted in Fig.~\ref{fig:windows}.  These window functions force proper normalized bins when we integrate over $l$ space. To factor in the area of each survey, we note that the ACBAR survey covers 600 sq. degrees and CBI covers 143 sq. degrees.

Fig.~\ref{fig:ratio} shows the diagonal ratio of the covariance matrix
together with non-Gaussian contributions, with the purely Gaussian
covariance matrix, for the ACBAR and CBI surveys.  We plot only the
diagonal, as these entries are several orders of magnitude larger than the
off-diagonal pieces. All contributions along the diagonal of the covariance leave
corrections no larger than on the order of $1\%$.

\begin{figure}
\centering
\includegraphics[scale=1.15]{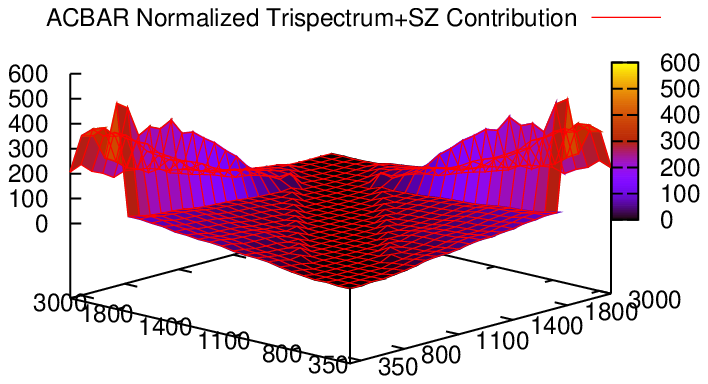}
\centering
\includegraphics[scale=1.15]{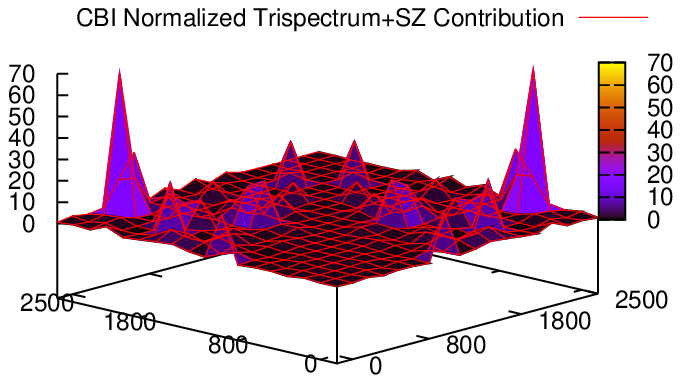}
\caption{The absolute values of the Gaussian covariance plus trispectrum matrix divided by the Gaussian covariance matrix alone. In the top figure we show corrections to the ACBAR data set, and in the lower figure to the CBI data set.
The trispectrum contribution to the cross-correlations is about 100 times larger than the Gaussian contribution.}
\label{fig:cbi_divide}  
\end{figure}

Cross-correlations become significant when the trispectrum is considered, since a fully Gaussian theory should not have any cross-correlation terms in the covariance of the temperature power spectrum. As seen in Fig.~\ref{fig:cbi_divide}, the non-Gaussian corrections to the cross-correlations lie on the order of a factor of 100 as they correct off-diagonal terms that otherwise are negligible.
However, as will be shown, despite the large corrections to the cross-correlations, the trispectrum effects on the cosmological parameters are negligible.

Having obtained corrections to the binned covariance matrix from
trispectrum effects, we now compute the impact of these on the cosmological parameter constraints.  The method we use in our analysis is based on the publicly available Markov Chain Monte Carlo (MCMC) package CosmoMC~\cite{Lewis:2002ah} with a convergence diagnostic done through the Gelman and Rubin statistics. We sample the following seven-dimensional set of cosmological parameters, adopting flat priors on them: the baryon and cold dark matter densities $\Omega_b$ and $\Omega_c$, the ratio of the sound horizon to the angular diameter distance at the decoupling, $\theta_s$, the scalar spectral index $n _s$, the overall normalization of the spectrum $A_s$ at $k=0.002$ Mpc$^{-1}$, the optical depth to reionization, $\tau$, and the weak lensing parameter $A_L$. We use the same window functions and band powers supplied from the considered
experiments (see Fig.~\ref{fig:windows}). We only use the band
powers for $l > 900$ for both ACBAR and CBI in our CosmoMC analysis,
and use WMAP data for calculations on larger angular scales. The non-Gaussian contributions
to the trispectrum are incorporated into the binned covariance matrix for use in CosmoMC.

\begin{figure}[!t]
\centering
\includegraphics[scale=0.5]{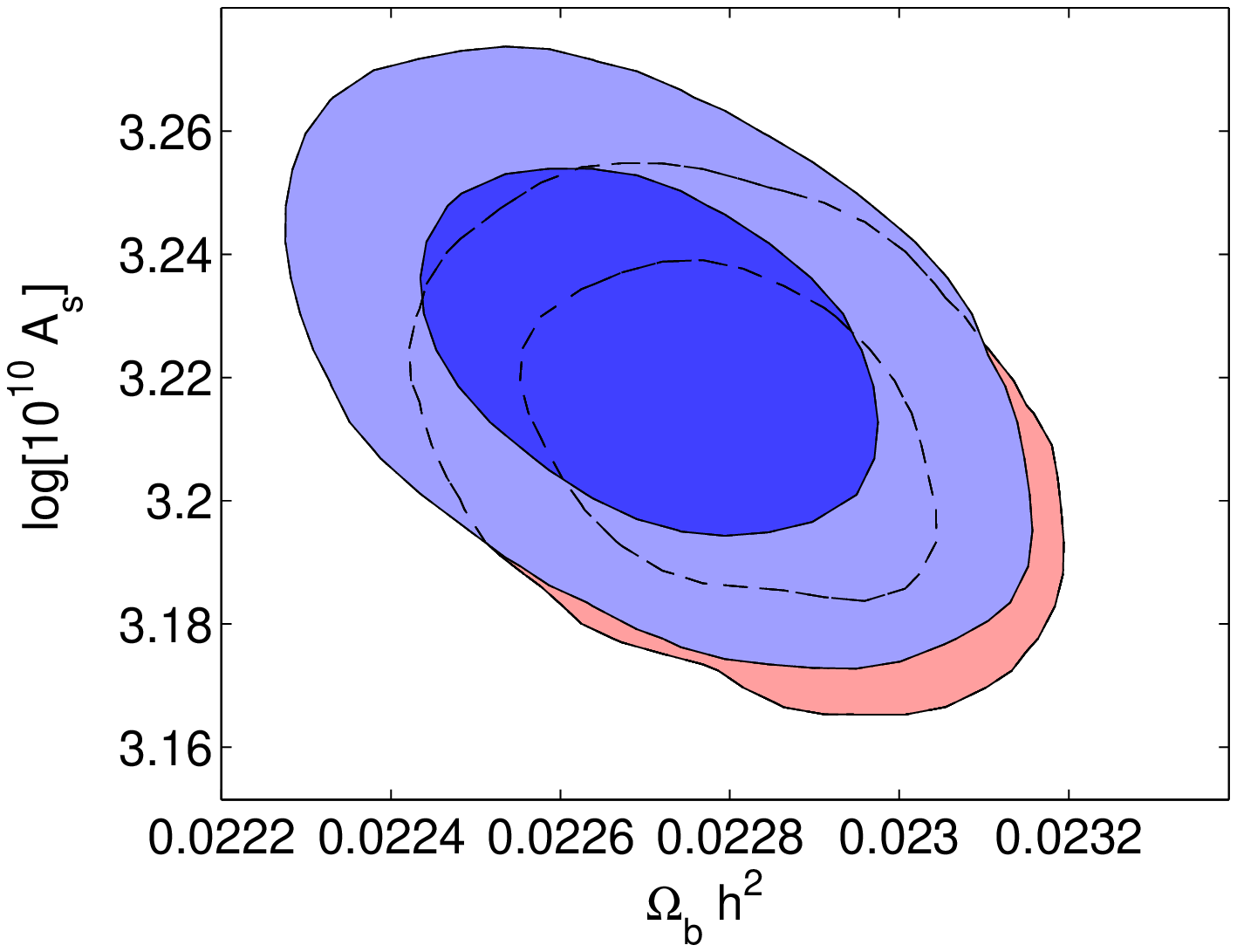}
\centering
\includegraphics[scale=0.5]{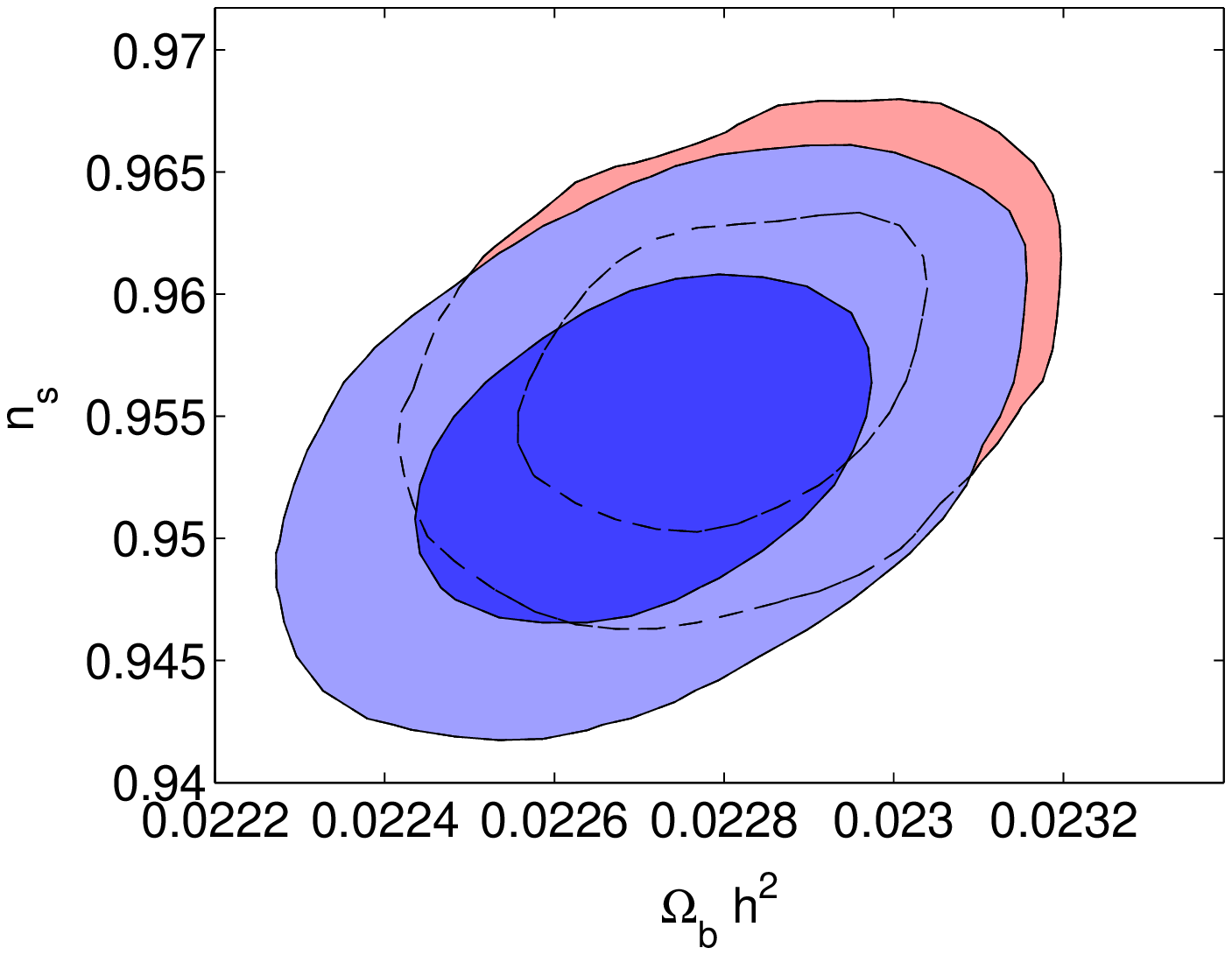}
\includegraphics[scale=0.5]{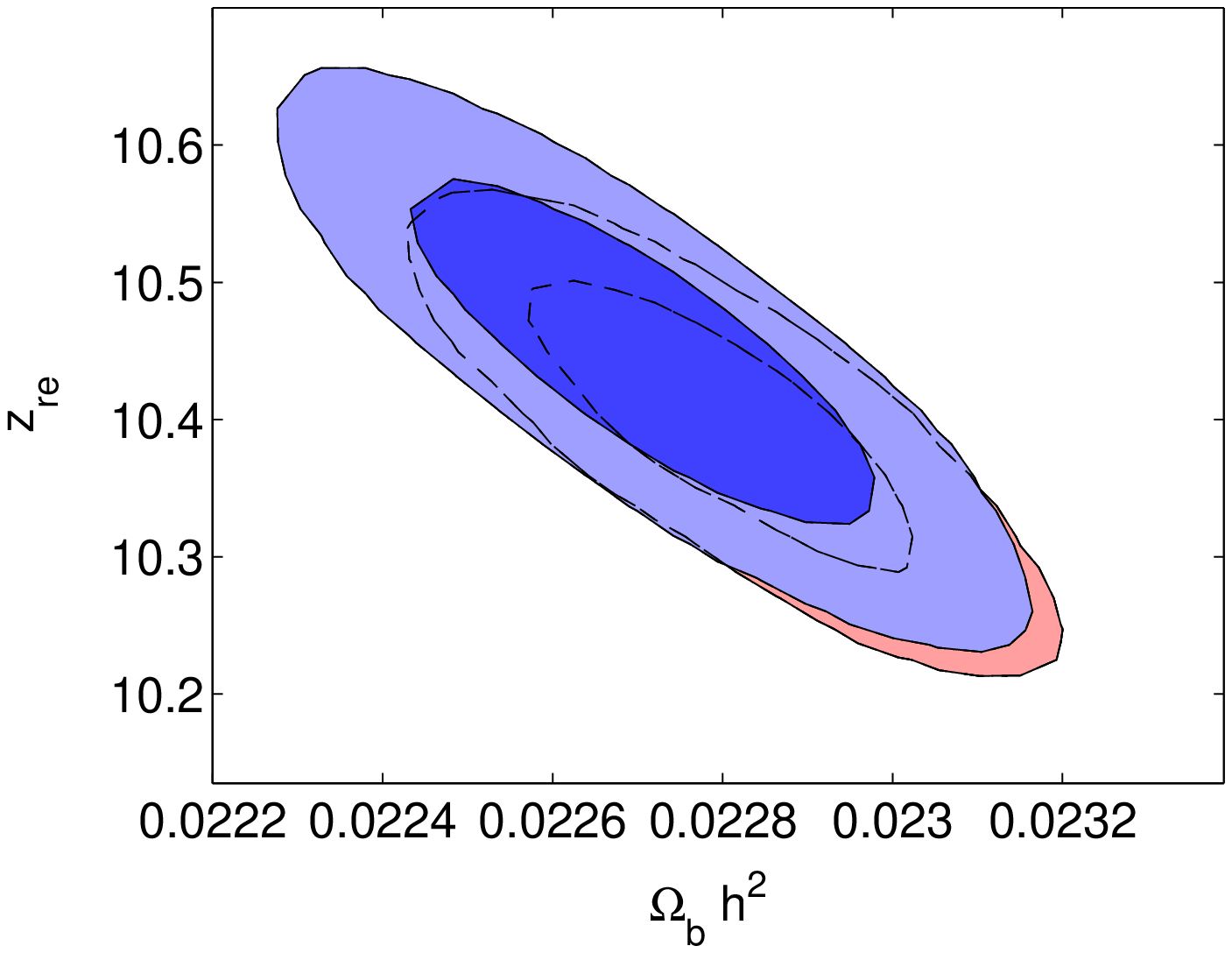}
\caption{The 68\% and 95\% confidence intervals in cosmological parameter constraints for the mock Planck run. These parameters are those with the largest shifts as seen in Table \ref{tab:planck} plotted against $\Omega_b h^2$. The red regions are those without the trispectrum, whereas the blue regions are those with the trispectrum.}
\label{fig:contours}  
\end{figure}

\begin{table*}
\centering
\space
\caption{
The first column represents the cosmological parameters considered. In the second column we quantify the parameter constraints without non-Gaussian effects. The cosmological parameters
determined from the combined ACBAR, CBI and WMAP five-year CMB data sets along with the non-Gaussian contributions from the trispectrum are shown in the third column.   The same data without the trispectrum but with the variation of $A_L$ is shown in the fifth column.  The columns denoted $\Delta \sigma$ represent the percent change of the 1$\sigma$ errror bars.   
}    
\begin{tabular}{|c||c||c||c||c||c|}
\hline\hline
Parameter & CMBData & CMBData+Tri.  & $\Delta \sigma / \sigma$ & CMBData+$A_{L}$  & $\Delta \sigma / \sigma$
 \\
\hline
$    \Omega_b h^2 $  &  $ 0.02284 \pm 0.00058 $  &  $ 0.02283 \pm
0.00059 $ & $ 2\% $ &  $ 0.02300 \pm
0.00063 $ & $ 9\% $ \\
$    \Omega_c h^2 $  &  $ 0.1103 \pm 0.0060 $  &  $ 0.1105 \pm 0.0061
$ & $ 2\% $ &  $ 0.1082 \pm 0.0061
$ & $ 2\% $ \\
$    \theta $  &  $ 1.0408 \pm 0.0026 $  &  $ 1.0410 \pm 0.0028 $ & $ 8\% $ &  $ 1.0413 \pm 0.0028 $ & $ 8\% $ \\
$    \tau $  &  $ 0.090 \pm 0.017 $  &  $ 0.089 \pm 0.017 $ & $ 0\% $ &    $ 0.090 \pm 0.017$ & $ 0\% $  \\
$    n_s $  &  $ 0.966 \pm 0.014 $  &  $ 0.966 \pm 0.015 $ & $ 7\% $ &  $ 0.968 \pm 0.015 $ & $ 7\% $ \\
$    A_{L} $  &1&1 &  -  &  $ 2.32 \pm 0.93 $ & -  \\
$    {\rm log}[10^{10} A_s] $  &  $ 3.180 \pm 0.047 $  &  $ 3.178 \pm 0.048
$ & $ 2\% $  &  $ 3.166 \pm 0.048
$ & $ 2\% $\\
$    \Omega_\Lambda $  &  $ 0.741 \pm 0.029 $  &  $ 0.740 \pm 0.029 $ &
$ 0\% $ &  $ 0.752 \pm 0.030 $ &
$ 3\% $  \\
$    {\rm Age/GYr} $  &  $ 13.67 \pm 0.13 $  &  $ 13.67 \pm 0.14 $ & $ 8\% $ &  $ 13.63 \pm 0.14 $ & $ 8\% $  \\
$    \Omega_m $  &  $ 0.259 \pm 0.029 $  &  $ 0.259 \pm 0.029 $ & $ 0\% $  &  $ 0.247 \pm 0.029 $ & $ 0\%$ \\
$    \sigma_8 $  &  $ 0.815 \pm 0.031 $  &  $ 0.801 \pm 0.033 $ & $ 7\% $ &  $ 0.789 \pm 0.035 $ & $ 13\% $\\
$    z_{re} $  &  $ 10.5 \pm 1.4 $  &  $ 10.4 \pm 1.4 $ & $ 0\% $ &  $ 10.4 \pm 1.4 $ & $ 0\% $ \\
$    H_0 $  &  $ 71.9 \pm 2.6 $  &  $ 71.8 \pm 2.7 $ & $ 4\% $ &  $ 73.1 \pm 2.8 $ & $ 8\% $ \\
\hline\hline
\end{tabular}
\label{tab:orig}
\end{table*}

\begin{table*}
\label{table:runTri}
\centering
\space
\caption{
In the second column, we quantify the cosmological parameter constraints incorporating the additional free
parameters $A_L$ and $dn_s/dlnk$. In the third column, we show the cosmological parameters determined from the combined ACBAR, CBI and WMAP 5-year CMB datasets with contributions from the trispectrum, along with the
additional free parameters $A_L$ and $dn_s/dlnk$.  
}.    
\begin{tabular}{|c||c||c||c|}
\hline\hline
Parameter & CMBData+$A_L$+Run & CMBData+$A_{L}$+Run+Tri.  &
$\Delta \sigma / \sigma$ \\
\hline
$    \Omega_b h^2 $  &  $ 0.02283 \pm 0.00080 $  &  $ 0.02269 \pm
0.00081 $ & $ 2\% $ \\
$    \Omega_c h^2 $  &  $ 0.1135 \pm 0.0089 $  &  $ 0.1136 \pm 0.0090
$ & $ 1\% $  \\
$    \theta $  &  $ 1.0410 \pm 0.0029 $  &  $ 1.0410 \pm 0.0030 $ & $
3\% $  \\
$    \tau $  &  $ 0.090 \pm 0.017 $  &  $ 0.090 \pm 0.018 $ & $ 6\% $  \\
$    n_s $  &  $ 0.997 \pm 0.048 $  &  $ 1.001 \pm 0.049 $ & $ 2\% $  \\
$    A_{L} $  &  $ 2.10 \pm 0.90$  &  $ 2.02 \pm 0.94 $ & $ 4\% $  \\
$    dn_s/dlnk $  &  $ -0.014 \pm 0.024 $  &  $ -0.017 \pm 0.025 $ & $
4\%$  \\
$    {\rm log}[10^{10} A_s] $  &  $ 3.167 \pm 0.049 $  &  $ 3.167 \pm 0.049
$ & $ 0\%$  \\
$    \Omega_\Lambda $  &  $ 0.726 \pm 0.047 $  &  $ 0.723 \pm 0.048 $ &
$ 2\% $ \\
$    {\rm Age/GYr} $  &  $ 13.67 \pm 0.16 $  &  $ 13.69 \pm 0.17 $ & $ 6\% $  \\
$    \Omega_m $  &  $ 0.275 \pm 0.029 $  &  $ 0.277 \pm 0.029 $ & $ 0\% $  \\
$    \sigma_8 $  &  $ 0.810 \pm 0.039 $  &  $ 0.807 \pm 0.041 $ & $ 5\%$ \\
$    z_{re} $  &  $ 10.8 \pm 1.5 $  &  $ 10.8 \pm 1.5 $ & $ 0\% $  \\
$    H_0 $  &  $ 71.0 \pm 4.0 $  &  $ 70.7 \pm 4.1 $ & $ 3\% $  \\
\hline \hline
\end{tabular}
\label{tab:origTri}
\end{table*}

\begin{table*}
\label{table:runTri}
\centering
\space
\caption{
In the second column, we show the cosmological parameter constraints from mock Planck data. In the third column, we quantify the non-Gaussianity induced shift of the best fit parameter values defined as the ratio of the parameter shift with the uncertainty. Simulation results combining the trispectrum with the Planck mock are shown in the third column.}.    
\begin{tabular}{|c||c||c||c||c|}
\hline\hline
Parameter & Planck & Planck + Tri.  & $\Delta p_i /\sigma$ &$\Delta \sigma$ \\
\hline
$    \Omega_b h^2 $  &  $ 0.02281 \pm 0.00016 $  &  $ 0.02271 \pm
0.00018 $ & $ -0.62 $ & 13\% \\
$    \Omega_c h^2 $  &  $ 0.1122 \pm 0.0016 $  &  $ 0.1132 \pm 0.0018
$ & $ 0.63 $ & 12\% \\
$    \theta $  &  $ 1.04095 \pm 0.00037 $  &  $ 1.04084 \pm 0.00044 $ & $
-0.30 $  & 19\%\\
$    n_s $  &  $ 0.9570 \pm 0.0042 $  &  $ 0.9538 \pm 0.0047 $ & $ -0.75 $  & 12\% \\
$    {\rm log}[10^{10} A_s] $  &  $ 3.211 \pm 0.017 $  &  $ 3.224 \pm 0.020
$ & $ 0.73 $  & 18\%\\
$    \Omega_\Lambda $  &  $ 0.7330 \pm 0.0086 $  &  $ 0.727 \pm 0.010 $ &
$ -0.67 $ & 14\% \\
$    {\rm Age/GYr} $  &  $ 13.686 \pm 0.028 $  &  $ 13.704 \pm 0.033 $ & $ 0.65 $ &  18\% \\
$    \Omega_m $  &  $ 0.2670 \pm 0.0086 $  &  $ 0.273 \pm 0.010 $ & $ 0.67 $ & 14\%  \\
$    \sigma_8 $  &  $ 0.8071 \pm 0.0068 $  &  $ 0.8113 \pm 0.0076 $ & $ -0.08 $ & 12\% \\
$    z_{re} $  &  $ 10.395 \pm 0.071 $  &  $ 10.447 \pm 0.081 $ & $ 0.74 $ & 14\%  \\
$    H_0 $  &  $ 71.12 \pm 0.76 $  &  $ 70.62 \pm 0.86 $ & $ -0.67 $  & 13\% \\
\hline \hline
\end{tabular}
\label{tab:planck}
\end{table*}

\begin{table*}
\label{table:runTri}
\centering
\space
\caption{
In the second column, we show the cosmological parameter constraints from mock CMBPol data. Simulation results combining the trispectrum with the CMBPol mock are shown in the third column.}.    
\begin{tabular}{|c||c||c||c||c|}
\hline\hline
Parameter & CMBPol & CMBPol + Tri.  & $\Delta p_i /\sigma$ &
$\Delta \sigma$ \\
\hline
$    \Omega_b h^2 $  &  $ 0.02292 \pm 0.00013 $  &  $ 0.02279 \pm
0.00016 $ & $ -1.01 $ & 23\% \\
$    \Omega_c h^2 $  &  $ 0.1111 \pm 0.0014 $  &  $ 0.1126 \pm 0.0017
$ & $ 1.09 $ &  21\% \\
$    \theta $  &  $ 1.04111 \pm 0.00028 $  &  $ 1.04110 \pm 0.00036 $ & $
-0.52 $ & 29\% \\
$    n_s $  &  $ 0.9603 \pm 0.0036 $  &  $ 0.9558 \pm 0.0044 $ & $ -1.23 $  & 22\%\\
$    {\rm log}[10^{10} A_s] $  &  $ 3.198 \pm 0.015 $  &  $ 3.216 \pm 0.018
$ & $ 1.19 $  & 20\% \\
$    \Omega_\Lambda $  &  $ 0.7392 \pm 0.0072 $  &  $ 0.7309 \pm 0.0091 $ &
$ -1.15 $ & 26\%\\
$    {\rm Age/GYr} $  &  $ 13.663 \pm 0.023 $  &  $ 13.689 \pm 0.030 $ & $ 1.11 $ & 30\% \\
$    \Omega_m $  &  $ 0.2608 \pm 0.0072 $  &  $ 0.2691 \pm 0.0091 $ & $ 1.09 $  & 26\%\\
$    \sigma_8 $  &  $ 0.8026 \pm 0.0057 $  &  $ 0.8088 \pm 0.0071 $ & $ -0.08 $ & 25\% \\
$    z_{re} $  &  $ 10.336 \pm 0.059 $  &  $ 10.410 \pm 0.073 $ & $ 1.25 $ & 24\%  \\
$    H_0 $  &  $ 71.69 \pm 0.65 $  &  $ 70.95 \pm 0.79 $ & $ -1.15 $ & 22\%  \\
\hline \hline
\end{tabular}
\label{tab:cmbpol}
\end{table*}

Table~\ref{tab:orig} shows the extent to which the trispectrum contributions affect the constraints on cosmological parameters for ACBAR and CBI data sets combined with WMAP five-year data. The trispectrum has a negligible impact on the cosmological parameters for these simulations.  For the cases where the weak lensing scaling parameter and the running of the spectral index are included the constraints on the cosmological parameters change by less than 10\%.

\subsection{The Weak Lensing Scaling Parameter}

Using a prior ACBAR dataset, it was found that the weak lensing scaling parameter $A_L$ is inconsistent with the expected value of unity at $2.5\sigma$~\cite{Calabrese:2008rt}. This led to an apparent revision by the ACBAR team and resulting in a recalibration of the data. This apparently removed the result first demonstrated by Calabrese et al. \cite{Calabrese:2008rt} and the ACBAR team's own revised analysis showed $A_L$ to be consistent with unity at a significance around 1.1$\sigma$ \cite{Reichardt:2008ay}.
However, given that this parameter is strongly sensitive to small angular scales where other secondary anisotropies contribute and non-Gaussianities become important, we have also chosen to include this parameter into our analysis. We have explicitly carried out two runs with this scaling parameter. For the first run, we incorporated  $A_L$ without any assumption of priors or the addition of other parameters, as shown in Table~\ref{tab:orig}. We find $A_L$ to be consistent with unity at $1.4\sigma$, for the same reasons that the ACBAR team found this parameter to
be consistent with the theoretical value in the fiducial cosmological model.

For the second simulation, in addition to $A_L$, we incorporated secondary non-Gaussianities and allowed the spectral index to run as this parameter is sensitive to physics on small scales. However, the constraint on the lensing amplitude remains unaffected. Nevertheless, it is possible that non-Gaussian effects are partly responsible for the inconsistency of the scaling parameter with unity, and future data sets with less noise on small angular scales are required to determine this.

\subsection{Mock Runs}

In order to understand the extent to which the trispectrum impacts the cosmological parameters in a prospective survey with less noise, we created TT mock datasets assuming the best fit WMAP5 cosmological parameters, with the optical depth fixed at $\tau = 0.087$, and with noise properties consistent with Planck's 143 GHz channel. Instead of considering cross-correlations, we only use the diagonal entries of the binned covariance matrix.

Table~\ref{tab:planck} shows the results of the Planck mock simulation. Since Planck has significantly less noise at high $l$, the trispectrum effects on the cosmological parameters is more apparent and the confidence regions of the most extreme changes are seen in Fig.~\ref{fig:contours}. Here, the 1$\sigma$ parameter constraints weaken by up to 20\%. Thus, the trispectrum has a more significant impact for measuring the constraints on the cosmological parameters in experiments with a noise-level comparable to that of Planck on small angular scales. 

Furthermore, in anticipation of CMBPol~\cite{Baumann:2008aq}, which should have noise an order of magnitude less than Planck, we have carried out a mock CMBPol-like run
with noise estimates consistent with the EPIC concept mission \cite{Bock}. 
The results of this run are seen in Table~\ref{tab:cmbpol}. For surveys with noise level comparable to that of CMBPol, non-Gaussian contributions in the trispectrum weaken the parameter constraints by up to 30\%. 

We have included in Table~\ref{tab:planck} and Table~\ref{tab:cmbpol} the parameter shifts induced in one realization of the mock CMB spectrum. Although the size of the parameter shifts extend up to 1.5$\sigma$, the shifts are not statistically significant as errors are correlated. We expect that for a set of realizations the shifts will cancel on average. 

\section{Conclusions}
\label{sec:summary}

Secondary non-Gaussianities in the trispectrum from weak lensing, the ISW effect, and the SZ effect have the potential to weaken the constraints on the cosmological parameters in experiments with high enough sensitivity to physics on small angular scales. While recent experiments such as ACBAR and CBI are too noisy for the trispectrum to become important at large multipoles, future surveys need to incorporate the trispectrum to avoid over-confident parameter estimation. The underlying cosmology constrained by Planck suffers a 20\% degradation when these secondary contributions to the temperature anisotropies are properly incorporated, and this rises to 30\% in CMBPol. Accounting for secondary non-Gaussianities in our analysis shows the weak lensing scaling parameter to be consistent with unity at $1.5\sigma$. 

\acknowledgments

We thank Erminia Calabrese, Eiichiro Komatsu, and 
Alessandro Melchiorri for helpful discussions. J.S. and S.J. acknowledge support from GAANN fellowships. This work was also supported by NSF AST-0645427.

\end{document}